\documentclass[a4paper,twoside]{article}

\usepackage{epsfig}
\usepackage{subcaption}
\usepackage{calc}
\usepackage{amssymb}
\usepackage{amstext}
\usepackage{amsmath}
\usepackage{amsthm}
\usepackage{multicol}
\usepackage{pslatex}
\usepackage{apalike}
\usepackage{algorithm2e}
\usepackage[bottom]{footmisc}
\usepackage{float}
\usepackage{algorithmic}
\usepackage{multirow}
\usepackage{placeins}
\usepackage{SCITEPRESS}     

\begin{document}

\title{Synergizing Data Imputation and Electronic Health Records for Advancing Prostate Cancer Research: Challenges, and Practical Applications.}

\author{\authorname{Abderrahim O. Batouche\sup{1,2,3,*}, Eugen Czeizler\sup{2,3,*}, Miika Koskinen\sup{4}, Tuomas Mirtti\sup{2,5} and Antti S. Rannikko\sup{2,6}}
\affiliation{\sup{1}Doctoral Programme in Computer Science, University of Helsinki, Helsinki, Finland}
\affiliation{\sup{2}Research Program in Systems Oncology, University of Helsinki, Helsinki, Finland}
\affiliation{\sup{3}ICAN-Digital Precision Cancer Medicine Flagship, Helsinki, Finland}
\affiliation{\sup{4}HUS Helsinki University Hospital, Helsinki, Finland}
\affiliation{\sup{5}Department of Pathology, University of Helsinki and Helsinki University Hospital, Helsinki, Finland}
\affiliation{\sup{6}Department of Urology, University of Helsinki and Helsinki University Hospital, Helsinki, Finland}
\affiliation{\sup{*}These authors contributed equally to this work.}
\email{\{abderrahim.batouche, eugen.czeizler\}@helsinki.fi, \{miika.koskinen, tuomas.mirtti, antti.rannikko\}@hus.fi}
}


\abstract{The presence of detailed clinical information in electronic health record (EHR) systems presents promising prospects for enhancing patient care through automated retrieval techniques. Nevertheless, it is widely acknowledged that accessing data within EHRs is hindered by various methodological challenges. Specifically, the clinical notes stored in EHRs are composed in a narrative form, making them prone to ambiguous formulations and highly unstructured data presentations, while structured reports commonly suffer from missing and/or erroneous data entries. This inherent complexity poses significant challenges when attempting automated large-scale medical knowledge extraction tasks, necessitating the application of advanced tools, such as natural language processing (NLP), as well as data audit techniques. This work aims to address these obstacles by creating and validating a novel pipeline designed to extract relevant data pertaining to prostate cancer patients. The objective is to exploit the inherent redundancies available within the integrated structured and unstructured data entries within EHRs in order to generate comprehensive and reliable medical databases, ready to be used in advanced research studies. Additionally, the study explores potential opportunities arising from these data, offering valuable prospects for advancing research in prostate cancer.}

\onecolumn \maketitle \normalsize \setcounter{footnote}{0} \vfill

\section{\uppercase{Introduction}}

Prostate cancer (PCa) is a prevalent disease known for its indolent nature, often characterised by slow development and protracted progression over time \cite{cancer_gov,cancer_org}. As such, one specific challenge in performing medical research pertaining to PCa is dealing with incomplete medical records and missing data, e.g., as a result of city relocation or disease follow-up across different health providers. This, in turn, can hinder the results of ongoing research studies analysing the effectiveness of diagnoses and various treatment planning approaches \cite{holmes_why_2021}. Ultimately, this can affect clinical decision-making and the patient's well-being.

To overcome the limitations of incomplete and/or erroneous data, Electronic Health Records (EHRs) mining has emerged as a crucial approach in medical research as well as within clinical practice\cite{yadav_mining_2018}. EHRs mining leverages advanced data analytic and artificial intelligence (AI) approaches to extract valuable insights from vast amounts of patient data\cite{ajmal_natural_2023,javaid_significance_2022}. By identifying patterns, trends, and risk factors associated with prostate cancer, EHRs mining facilitates the early detection of advanced diseases and the personalisation of treatment strategies \cite{knighton_using_2016,seneviratne_identifying_2018,henkel_initial_2022}. However, challenges such as missing data and data security must be addressed to ensure patient information remains complete, confidential, and secure. Additionally, the lack of interoperability between different EHR systems poses hurdles in data sharing and aggregation, limiting the full potential of mining for both prostate cancer research, as well as for general improvement of patient care \cite{de_la_torre-diez_ehr_2013}. Overcoming these issues and promoting standardised data collection practices and protocols will be pivotal in advancing the field of PCa treatment through EHRs mining\cite{10114915}, as well as the overall medical research in general.

In our work, we have designed and developed a data preprocessing pipeline that can leverage routinely collected information from our EHRs (HUS Datalake \cite{oscar_bruck_hus_2023,tietoevrycom_hus_nodate,Misukka2022}) to efficiently and accurately retrieve and consolidate clinicians’ work on PCa treatment analysis. Using Microsoft Azure machine learning studio and batches from HUS datalake that are available at the HUS Acamedic environment (a secure scalable data analytics platform developed for medical research \cite{noauthor_hus_nodate}), we developed an EHR mining pipeline using Python libraries to read, process, and provide curated data for further research applications.

One of the key clinical inputs exhibiting missing entries within the EHR of a significant number of PCa patients is the occurrence of curative treatment, i.e., radical prostatectomy (RP) or radiation treatment (RT). Since imputation of such missing data is inevitable, we had to use a different approach to uncover these lost data entries. Using routinely collected values of the prostate-specific antigen (PSA) lab measurements, we were able to successfully identify and even classify curative PCa treatments. To our knowledge, this is the first attempt to approach the inference of EHR missing treatment records through PSA time series data. Our approach enabled us to enhance our EHR by incorporating approximately 2.8 thousand new curative treatment events, marking a notable 27\% growth compared to the treatment events available beforehand. The explanation for this relatively large increase is multi-folded. Some patients might have been treated outside the (Helsinki and Uusima) district unit whose database our study is based upon. Others might have been treated within private practice units, which again are not covered by our database. Finally, we can assume that a proportion of these missing treatment events are due to human error in correctly recording them within the EHR.

Another key clinical information (as well as key surrogate measurement within medical research analysis) which is most of the times not directly recorded within EHRs, either in structured or non-structured format, is the time instant when PCa patients are classified as having a biochemical recurrence (BCR). After primary cancer treatment, BCR is achieved when the PSA level in the blood surpasses a certain threshold, thus indicating that the disease may be returning or progressing. Thus, BCR status is an important indicator both clinically, as it signs that further monitoring or treatment may be needed to manage the condition \cite{bcr_ref_JCO,bcr_a_riview}, and from a (medical) data analysis perspective, as it is a surrogate for PCa mortality \cite{why_bcr_pca_surrogate,bcr_a_riview}. By following the PSA measurements as well as all EHR-available PCa treatment records we were able to effectively determine (and report) the status and timing of BCR for all PCa patients. 

\section{\uppercase{Methods}}

\subsection{Data source}
\begin{figure}[!ht]
    \centering
    \includegraphics[width=1\linewidth]{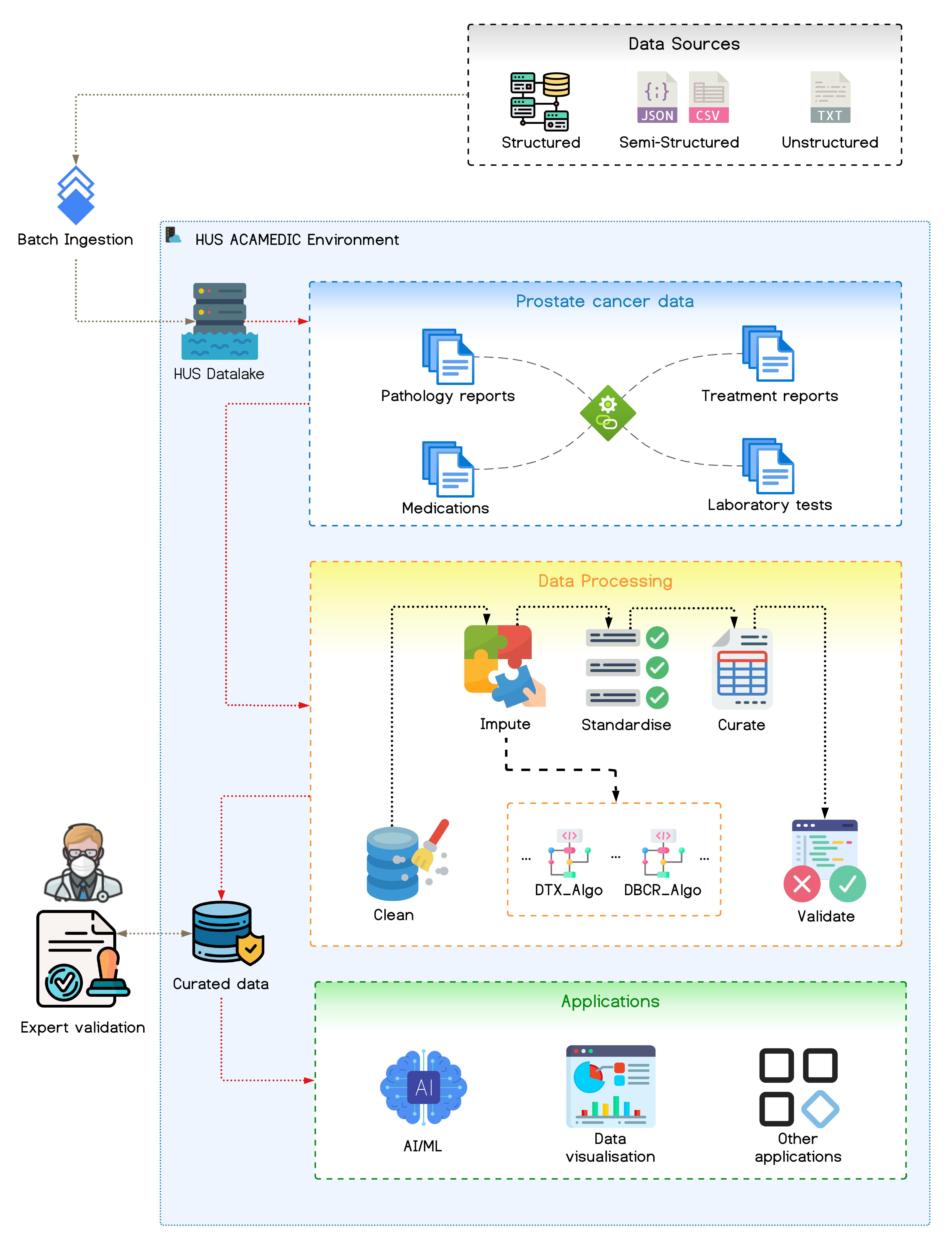}
    \caption{Data preprocessing pipeline for Prostate Cancer research data.}
    \label{fig:main}
\end{figure}

Our pipeline starts by identifying patients of interest within a large academic EHR system (Figure \ref{fig:main}). We used the Finnprostate dataset, which is a large patient registry study combining Finnish national healthcare data with local hospital data (n=700,000) of men suspected of having PCa (PSA measured) or diagnosed with PCa. From Finnprostate, we gathered a HUS (Hospital District of Helsinki and Uusimaa) sub-cohort of men (n=326,796) having comprehensive patient information regarding out-patient clinic and hospital visits as well as data regarding laboratory tests, medication prescriptions, radiological, pathological, and surgical reports, as well as comorbidities covering the years 1993 to 2019. The above data is embedded within the regional HUS Acamedic datalake. 

Medical research commonly encounters missing data. Despite this prevalence, it is nowadays generally accepted to perform various data analysis tasks on partially incomplete records, as long as the missing values are not substantial, and the analysis methods themselves can cope with specific uncertainties. Moreover, the use of advanced imputation techniques such as maximum likelihood \cite{max_liklihood}, multiple imputation\cite{multiple_imputation}, or Bayesian methods \cite{bayesian} have a good track record in addressing many of the missing data entries. However, certain complex missing data records, such as the moment and type of a deployed treatment, or the first diagnostic biopsy of a tumour and its aggressiveness, are very hard to be addressed by any of the available computational imputation methods.

In our data processing work (yellow box, Figure \ref{fig:main}), imputation was reinforced with customised algorithms that rely on clinical guidelines, experts’ interpretations, as well as the intrinsic information redundancy available within EHR, in order to retrieve the missing data. All created algorithms are described in Table \ref{algo-summary-table}.

\subsection{Missing curative treatments detection}
The Treatment Detection Algorithm (DTX\_algo) plays a pivotal role in enhancing our data quality by identifying and incorporating missing curative treatment records (in Algorithm \ref{algo_dtx}). The algorithm takes all patient's $data$ as an input and returns a list of missing curative treatments. 

The Significant PSA Drop Algorithm (SIGDROP) constitutes the initial phase of DTX, meticulously tracking a patient's PSA values subsequent to their diagnostic biopsy (Algorithm \ref{algo_sigdrop}). The algorithm takes PSA measurements of patient $i$, and returns, if any:
\begin{itemize}
    \item $drop\_date$: The date of the PSA drop, which is the highest (maximum) point from where a significant PSA drop starts; is subsequently considered as a treatment date.
    \item  $nadir\_date$: The date of the PSA nadir, which is the lowest (minimum) point to where the significant drop reached.
    \item  $PSA_{min}$: The minimum values (at the time $nadir\_date$); this value is used to classify the drop into RP or RT.
\end{itemize}

The algorithm's operation commences with the pursuit of the maximum PSA value ($PSA_{max}$, line 3-4), followed by an endeavour to identify the minimum value within the ensuing $\delta \leq$ 12-month period (lines 5-32). Upon successful identification of a decreasing value, at lines 15-16, the algorithm calculates $\alpha$, which is the drop percentage that undergoes rigorous testing to ascertain its adherence to predetermined significance conditions (line 17). This process is indispensable in establishing the genuineness of the observed drop and confirming its clinical significance.
\begin{table*}[h]
    \centering
    \caption{Summary of Algorithms}
    \label{algo-summary-table}
    \scriptsize
    \begin{tabular}{c c c c p{5cm} }
        \hline
        \textbf{Algorithm Name} & \textbf{Input} & \textbf{Output} & \textbf{Complexity} & \textbf{Short Description} \\
        \hline
        SIGDROP & $PSA_i$ & $drop\_date$, $nadir\_date$, $PSA_{min}$ & $O(M)$ & Detects significant PSA drop and related dates. \\
        \hline
        DTX & $PATIENTS\_LIST$ & $L$ & $O(M*N)$ & Detects missing treatments based on PSA data. \\
        \hline
        CRT & $p_i$ & $d_{m1}$ & $O(1)$ & Detects Clinical Relapse after RT \\
        \hline
        CRP & $p_i$ & $d_{m1}$ & $O(1)$ & Detects Clinical Relapse after RP \\
        \hline
        PRP & $p_i$ & $d_{m2}$ & $O(N)$ & Detects PSA Relapse after RP \\
        \hline
        PRT & $p_i$ & $d_{m2}$ & $O(N)$ & Detects PSA Relapse after RT \\
        \hline
        DBCR & $TREATED\_PATIENTS$ & $L_{bcr}$ & $O(M*N)$ & Main algorithm for BCR detection. \\
        \hline
        \multicolumn{4}{l}{- RP=Radical prostatectomy, RT=Radiation therapy, BCR=Biochemical recurrence.} \\
        \multicolumn{4}{l}{- In M*N  M is the number of PSA measurments and N is the number of patients} \\
    \end{tabular}
\end{table*}

\begin{algorithm}[h]
 \caption{SIGDROP - Significant PSA drop detection}
 \label{algo_sigdrop}
 \begin{algorithmic}[1]
 \renewcommand{\algorithmicrequire}{\textbf{Input:}}
 \renewcommand{\algorithmicensure}{\textbf{Output:}}
 
 \REQUIRE $PSA_i$
 \ENSURE  $drop\_date$, $nadir\_date$, $PSA_{min}$
 
  \STATE $M \leftarrow size(PSA_i)$

  \IF {$M \geq 0$}
    \STATE $PSA_{max} \leftarrow PSA_i[1]$
    \STATE $date\_PSA_{max} \leftarrow getDate(PSA_{max})$

    \FOR{$j = 1$ to $M-1$}
        \STATE $e \leftarrow PSA_i[j] - PSA_i[j+1]$
        \STATE $\delta \leftarrow date\_PSA_{next} - date\_PSA_{max} $
        \IF{$e \leq 0$}
            \STATE $date\_PSA_{next} \leftarrow getDate(PSA_i[j+1])$
            \IF{($PSA_{max} < PSA_i[j+1]$) \textbf{or} $\delta > 12m$}
                \STATE $PSA_{max} \leftarrow PSA_i[j+1]$
                \STATE $date\_PSA_{max} \leftarrow getDate(PSA_{max})$
            \ENDIF
        \ELSE
            \STATE $\beta \leftarrow  PSA_{max} - PSA_i[j+1] $
            \STATE $\alpha \leftarrow  \frac{\beta}{PSA_{max}}$
            
            \IF{($\alpha \geq 0.75$ \textbf{and} $\beta \geq 3 $) \textbf{or} ($\alpha \geq 0.5$ \textbf{and} $\beta \geq 4 $)}
                \STATE $PSA_{min} \leftarrow PSA_i[j+1]$
            \ELSE
                \IF{$\delta > 12m$}
                    \STATE $PSA_{max} \leftarrow PSA_i[j+1]$
                    \STATE $date\_PSA_{max} \leftarrow getDate(PSA_{max})$ 
                \ELSE
                    \STATE $\gamma \leftarrow date\_PSA[j+2] - date\_PSA_{max} $
                    \IF{$j+2 \leq M$ \textbf{and} $\gamma > 12$}
                        \STATE $PSA_{max} \leftarrow PSA_i[j+1]$
                        \STATE $date\_PSA_{max} \leftarrow getDate(PSA_{max})$
                    \ENDIF
                \ENDIF
            \ENDIF
        
        \ENDIF
    \ENDFOR
    
  \ENDIF

 \IF{$PSA_{min}$ \textbf{exists}}
 \STATE $drop\_date \leftarrow get\_date(PSA_{max})$
 \STATE $nadir\_date \leftarrow get\_date(PSA_{min})$
 \RETURN $drop\_date, nadir\_date ,PSA_{min}$
 \ENDIF
 \RETURN $NULL$
 
 \end{algorithmic}
\end{algorithm}

Having validated the drop as significant, and (line 6) with no EHR-recorded curative treatment between the date of drop ($d_{max}$) and the date of the nadir ($d_{min}$), DTX proceeds to collate all such identified drops, systematically categorising them into two distinct treatment modalities: radiation therapy (RT) and radical prostatectomy (RP) (Algorithm 2 line 7-10). This classification not only facilitates comprehensive treatment record augmentation but also provides valuable missing insights into the patient's therapeutic journey.

\subsection{Biochemical recurrence detection}
Biochemical recurrence (BCR) serves as a crucial indicator for PCa mortality. However, its availability in EHRs is not always guaranteed. In such cases, various methods can be employed to retrieve and impute this information. Our Detect Biochemical Recurrence (DBCR\_algo) Algorithm is specifically designed to analyse data from treated patients, identifying potential relapses and categorising patients as either having experienced a BCR or not (Algorithm \ref{algo_dbcr_main}). To achieve this outcome, DBCR utilises four (04) distinct functions, each tailored to a specific task.

Clinical guidelines governing PSA relapse are stringent and clearly defined \cite{van_den_broeck_biochemical_2020}, and these guidelines are meticulously integrated into the PRP and PRT functions (Algorithms \ref{algo_prp}-\ref{algo_prt}). 
\begin{itemize}
    \item PSA-based relapse after radical prostatectomy $PRP(p_i)$: this function uses the European Association of Urology (EAU) guidelines \cite{van_den_broeck_biochemical_2020} to detect whether a PSA-based relapse occurred after radical prostatectomy. If an ultrasensitive PSA \cite{shen_ultrasensitive_2005} measurement $psa_j$ was taken for patient $p_i$ then we take this into consideration to define the maximum threshold (lines 3-7). 
    \item PSA-based relapse after radiation therapy $PRT(p_i)$: this function is also using the EAU guidelines \cite{van_den_broeck_biochemical_2020} to detect whether a PSA-based relapse occurred after radiation therapy. The algorithm searches for the first increase of 2 PSA units from a nadir value.
\end{itemize}

\begin{figure}[!h]
    \centering
    \includegraphics[width=0.85\linewidth]{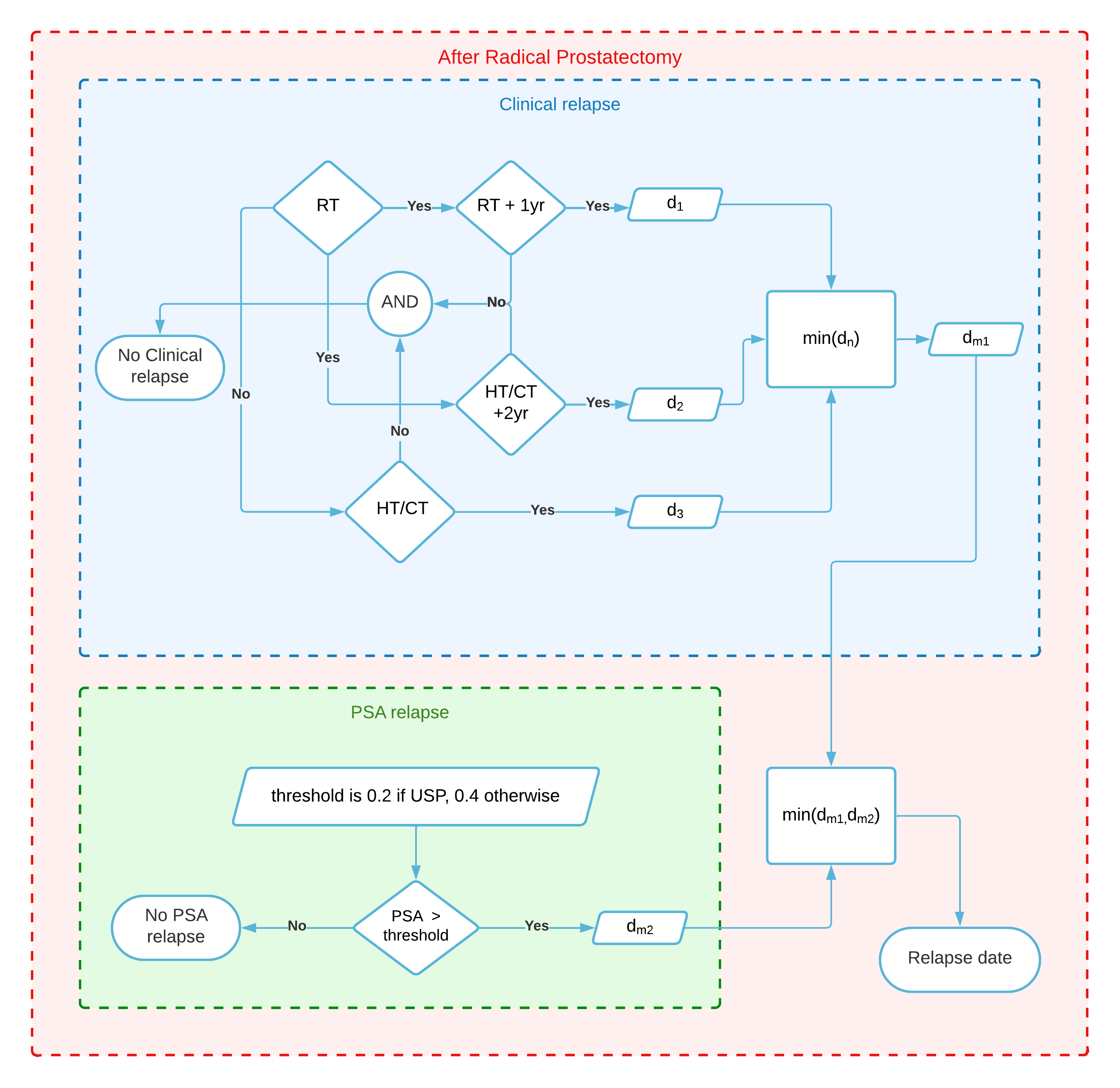}
    \caption{BCR definition after radical prostatectomy.}
    \label{fig:bcr_rp}
\end{figure}

\begin{figure}[!h]
    \centering
    \includegraphics[width=0.85\linewidth]{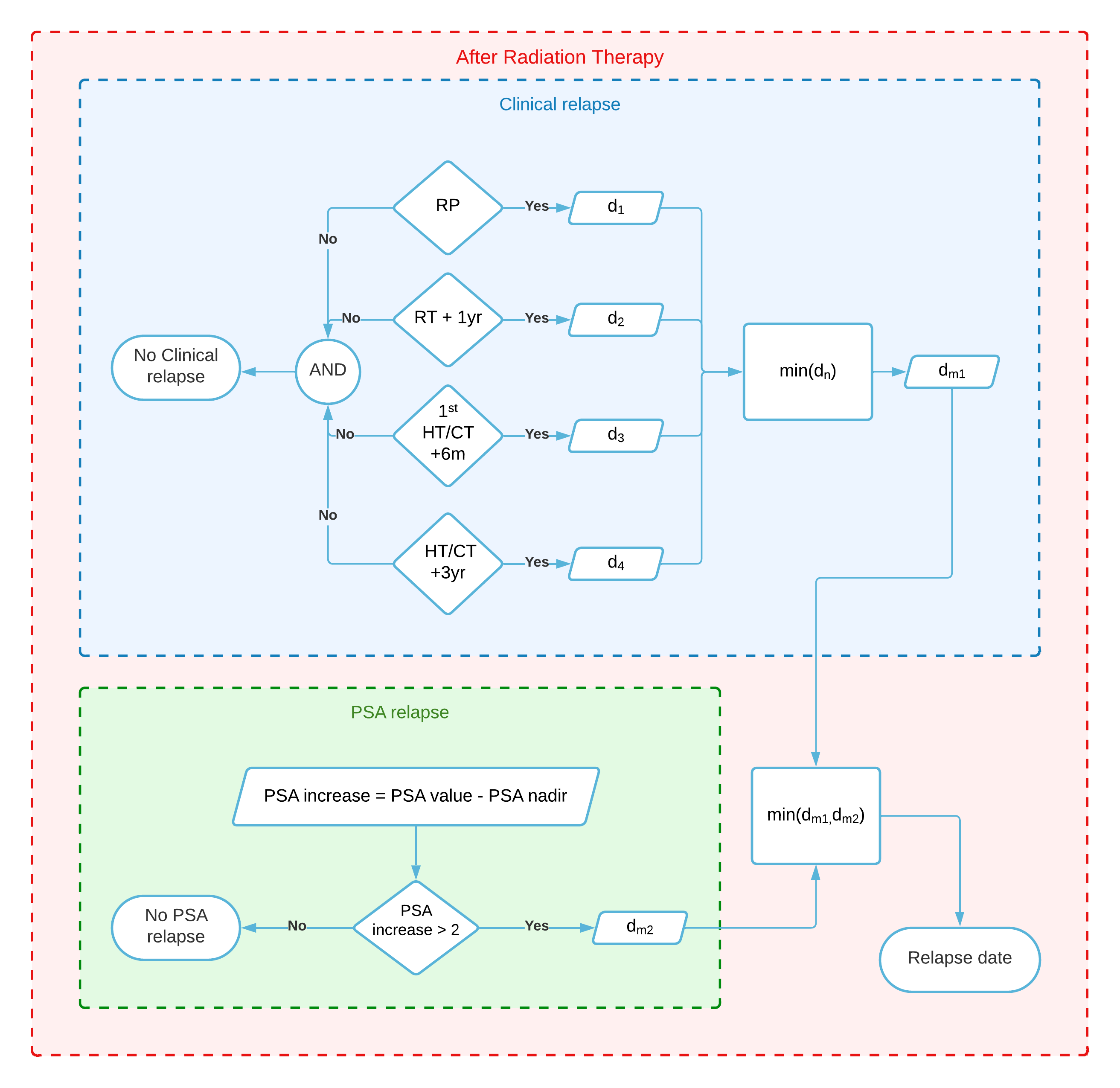}
    \caption{BCR definition after radiation therapy.}
    \label{fig:bcr_rt}
\end{figure}

Going beyond this, our novel BCR detection method is not solely reliant on PSA relapse; instead, it incorporates expert knowledge and translates it into a new tool for detecting BCR based on secondary treatments (Figures \ref{fig:bcr_rp} and \ref{fig:bcr_rt}). The CRP and CRT functions (Algorithms \ref{algo_crp}-\ref{algo_crt}) have been developed to identify possible relapses that may have been missed (after an RP or an RT primary treatment, respectively) either due to the absence of PSA tests or because the curating doctor decided on a secondary treatment before the PSA value has actually crossed the EAU-guideline threshold. The exact approaches used to define clinical relapse after RP and RT primary treatments are described in Figure  \ref{fig:bcr_rp} and Figure \ref{fig:bcr_rt}, respectively.


\begin{algorithm}[!ht]
 \centering
 \caption{CRP - Clinical Relapse after RP}
 \label{algo_crp}
 \begin{algorithmic}[1]
 \renewcommand{\algorithmicrequire}{\textbf{Input:}}
 \renewcommand{\algorithmicensure}{\textbf{Output:}}
 \renewcommand{\algorithmiccomment}[1]{\hfill$\triangleright$\textit{#1}}
 
 \REQUIRE $p_i$
 \ENSURE  $d_{m1}$
 
  \STATE $L \leftarrow [\;]$

    \IF{$lastRTDate(p_i) > firstRPDate(p_i)$}
        \IF{$lastRTDate(p_i) - firstRPDate(p_i) > 1yr$}
            \STATE $L \leftarrow L + firstRTDateAfterOneYear(p_i)$
        \ENDIF
        \IF{$hasHTCT(p_i) $ \textbf{and} $lastHTCTDate(p_i) > firstRPDate(p_i)$ }
            \IF{$lastHTCTDate(p_i)-firstRPDate(p_i) \geq 2yr$}
                \STATE $L \leftarrow L + firstHTCTDateAfterOneYear(p_i)$
            \ENDIF
        \ENDIF
    \ELSE
        \IF{$hasHTCT(p_i)$ \textbf{and} $lastHTCTDate(p_i) > firstRPDate(p_i)$}
            \STATE $L \leftarrow L + firstHTCTDateAfterRp(p_i)$
        \ENDIF
    \ENDIF

   \STATE $d_{m1} \leftarrow getMin(L)$
 \RETURN $d_{m1}$
 \end{algorithmic}
\end{algorithm}

\begin{algorithm}[!ht]
 \caption{CRT - Clinical Relapse after RT}
 \label{algo_crt}
 \begin{algorithmic}[1]
 \renewcommand{\algorithmicrequire}{\textbf{Input:}}
 \renewcommand{\algorithmicensure}{\textbf{Output:}}
 \renewcommand{\algorithmiccomment}[1]{\hfill$\triangleright$\textit{#1}}
 
 \REQUIRE $p_i$
 \ENSURE  $d_{m1}$
 
  \STATE $L \leftarrow [\;]$
   
   \IF{$hasRP(p_i)$ \textbf{and} $lastRPDate(p_i) > firstRTDate(p_i)$}    
        \STATE $L \leftarrow L + firstRpDateAfterRt(p_i)$
   \ENDIF

   \IF{$hasSecondRT(p_i)$ \textbf{and} $secondRTDate(p_i) - firstRTDate(p_i) > 1yr$}       
        \STATE $L \leftarrow L + secondRTDate(p_i)$
   \ENDIF

   \IF{$hasHTCT(p_i)$ \textbf{and} $firstHTCTDate(p_i) - firstRTDate(p_i) \geq 6m$}       
        \STATE $L \leftarrow L + firstHTCTDate(p_i)$
   \ENDIF

   \IF{$hasHTCT(p_i)$ \textbf{and} $firstHTCTDate(p_i) - firstRTDate(p_i) > 3yr$}       
        \STATE $L \leftarrow L + firstHTCTDate(p_i)$
   \ENDIF
   
   \STATE $d_{m1} \leftarrow getMin(L)$
 \RETURN $d_{m1}$
 
 \end{algorithmic}
\end{algorithm}

The DBCR Algorithm then uses all the outputs of the above functions, namely the dates ($d_1, d_2. d_3, d_4$) of possible BCR occurences, and selects the earliest date (if exists) as the date of biochemical recurrence for patient $p_i$ (Algorithm \ref{algo_dbcr_main} lines 7-10).

\subsection{Evaluation}
Retrieving missing data is of utmost importance in the pre-processing of EHR data for critical and sensitive applications. Additionally, assessing the quality of imputed data holds significant value as it provides insights into the effectiveness of the methods and algorithms employed. In our study, data evaluation involves a two-tier validation process. The first level (a.k.a. 'step-1' evaluation) employs automated tests, where we verify the accuracy of our algorithms by taking records without missing treatment data, applying the imputation algorithm, and subsequently scrutinising the outcomes. The second level (a.k.a. the 'step-2' evaluation) entails expert validation, wherein a random selection of imputed data is manually inspected by domain experts, ensuring its correctness.

\section{\uppercase{Results}}
\subsection{Curated database}
The initial phase of this work was to explore the HUS datalake and extract the most accurate and comprehensive data suitable for subsequent medical research applications. As a result, we successfully created a structured and curated database that contains crucial patient information, as defined in Table \ref{table_1}.

\begin{table}[H]
    \centering
    \caption{The curated data tables}
    \setlength\tabcolsep{0pt}
    \scriptsize
  \label{table_1}
    \begin{tabular*}{\columnwidth}{@{\extracolsep{\fill}} c c c }

     Data & Number of rows (\%) & Number of Patients (\%)\\
     \hline
     T1: Pathology        & 23,393    & 12,277\\
     \hspace{3mm} GG1 & 6618 (28) & 3652 (30)\\
     \hspace{3mm} GG2 & 6383 (27) & 3313 (27)\\
     \hspace{3mm} GG3 & 4747 (20) & 2503 (20)\\
     \hspace{3mm} GG4 & 2310 (10) & 1195 (10)\\
     \hspace{3mm} GG5 & 3335 (14) & 1614 (13)\\
     \hline
     T2: Treatment       & 40,369      & 9800 \\
     \hspace{3mm} RP & 2743   (7)  & 2742  \\
     \hspace{3mm} RT & 18,254 (45) & 7248  \\
     \hspace{3mm} HT & 15,804 (39) & 4088  \\
     \hspace{3mm} CT & 3568   (9)  &  514  \\
     \hline
     T3: PSA & 1,424,440 & 238,399\\
     \hline
     T4: MRI & 20,103    & 15,807\\
     \hline
     T5: Medications & 13,837,600 & 290,055\\
     \hline
    \multicolumn{3}{l}{GG1--GG5 = Gleason grade group 1--5 (associated to each pathological entry) } \\
    \multicolumn{3}{l}{RP=Radical prostatectomy, RT=Radiation therapy, } \\
    \multicolumn{3}{l}{HT=Hormonal therapy, CT=Chemotherapy.}
    \end{tabular*}
\end{table}

\subsection{Treatments data}
\begin{table*}[hbt!]
    \centering
    \caption{Evaluation of DTX algorithm performance}\label{table_evaluation}
    \scriptsize
    \begin{tabular}{c c c ccc c}
        
        - &Available CTx &Estimated CTx &\multicolumn{3}{c }{Correct estimated CTx} & New estimated CTx \\
        \hline
        - &DB &DTX &DTX $\cap$ DB &True-Class &False-Class & DTX $\setminus$ DB  \\
        \hline
        PID    &7563 &9725 &6962 (92\%) &6294 (90\%) &668 (10\%) &2763 (+27\%) \\
        PID-RP &2495 &2722 &2233 (90\%) &1929 (86\%) &304 (14\%) &0489  (+16\%) \\
        PID-RT &5068 &7003 &4729 (93\%) &4365 (92\%) &364  (08\%) &2274 (+31\%) \\
        \hline
    \end{tabular}
\end{table*}
Following the data curation and structuring, we have implemented the DTX algorithm in order to detect and impute the missing curative treatment data. As a result, our database now incorporates n=2763 new PCa-related treatment records, representing a 27\% increase compared to the original data found in the HUS datalake. The number of patients with RP has increased by 16\% (n=489), while the number of those with RT has increased by 31\% (n=2274).

In Table \ref{table_evaluation} we present the results of 'step-1' DTX performance evaluation, i.e., estimated vs. known (EHR-available) treatment records. We record an imputation performance of 92\% (n=6962) correct estimated curative treatments, i.e., treatments estimated using the DTX algorithm that are also found in the existing database. Out of these, 90\% (n=6294) are correctly classified as RP or RT, whereas 10\% (n=669) are wrongly classified. RP classification was 86\% correct, whereas RT classification reached 92\%.

The 'step-2' evaluation of the DTX algorithm was performed vs. manual validation by domain experts, where the experts were using the entire collection of unstructured reports associated with the test subjects in order to uncover their treatment history. The 'step-2' evaluation started by sampling 40 random patients, i.e., 20 random RP + 20 random RT, that were detected by the algorithm as having curative treatments (CTx), however this treatment did not appear within the EHR (DTX $\setminus$ DB in Table \ref{table_evaluation}). The results of this manual validation are summarised in Table \ref{table_manual_validation}. Only one patient from the RP group was unverifiable (no data = treatment cannot be confirmed), while five RT patients had the same situation. In addition, 95\% RP patients were confirmed to have a curative PCa treatment, and 60\% RT patients were confirmed. In total 79\% of the sampled patients (whose treatments were not recorded within EHR) were confirmed to have PCa curative treatment.

\begin{table}[hbt!]
    \centering
    \caption{Manual validation for DTX algorithm performance}\label{table_manual_validation}
    \setlength\tabcolsep{0pt}
    \scriptsize
    \begin{tabular*}{\columnwidth}{@{\extracolsep{\fill}}c c c c c c}
        
        - &Sample &Unverifiable &Verifiable &True CTx &All True CTx  \\
        \hline
        PID-RP &20 &01 &19 (95\%) &18 (95\%) & \multirow{2}{*}{27 (79\%)}  \\
        PID-RT &20 &05 &15 (75\%) &09 (60\%)   \\
        \hline
    \end{tabular*}
\end{table}

\subsection{BCR data}
\begin{figure}[!h]
    \centering
    \includegraphics[width=1\linewidth]{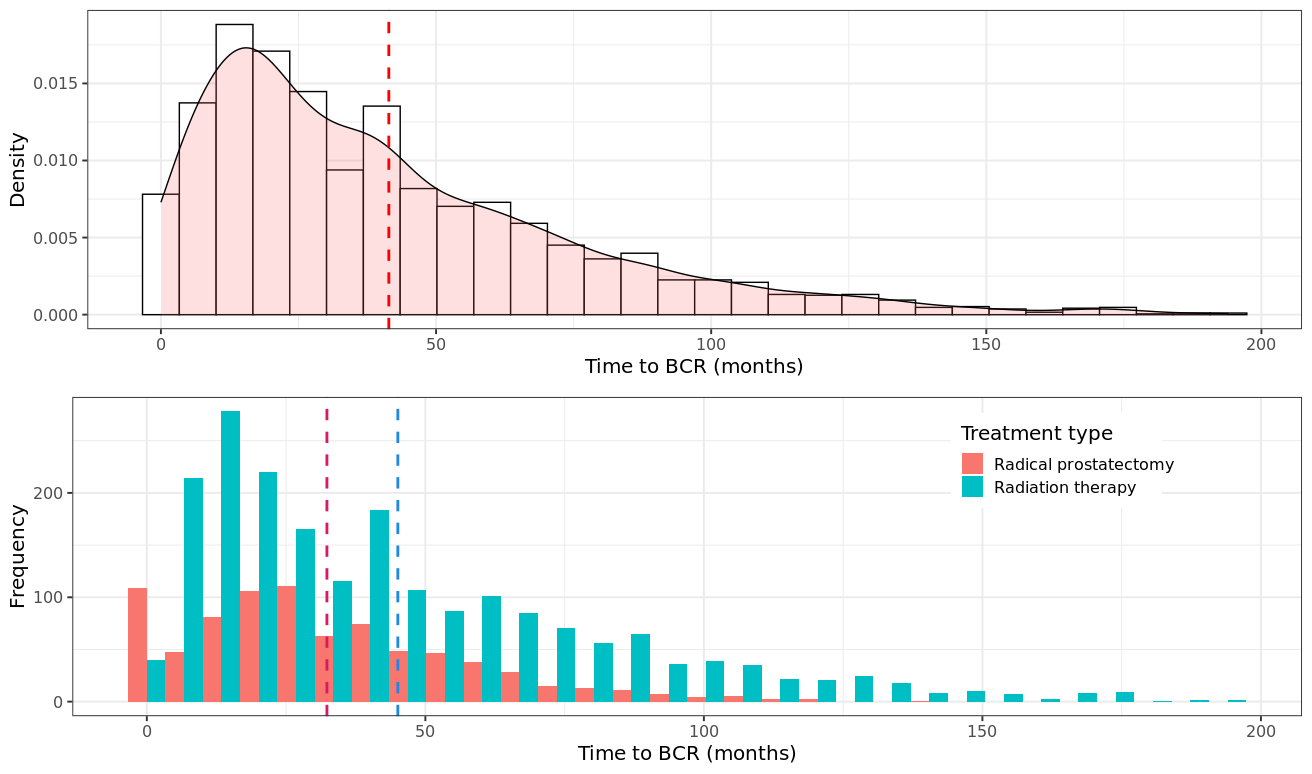}
    \caption{BCR detected data: The distribution of time to relapse.}
    \label{fig:bcr_dist}
\end{figure}

\begin{figure}[!h]
    \centering
    \includegraphics[width=1\linewidth]{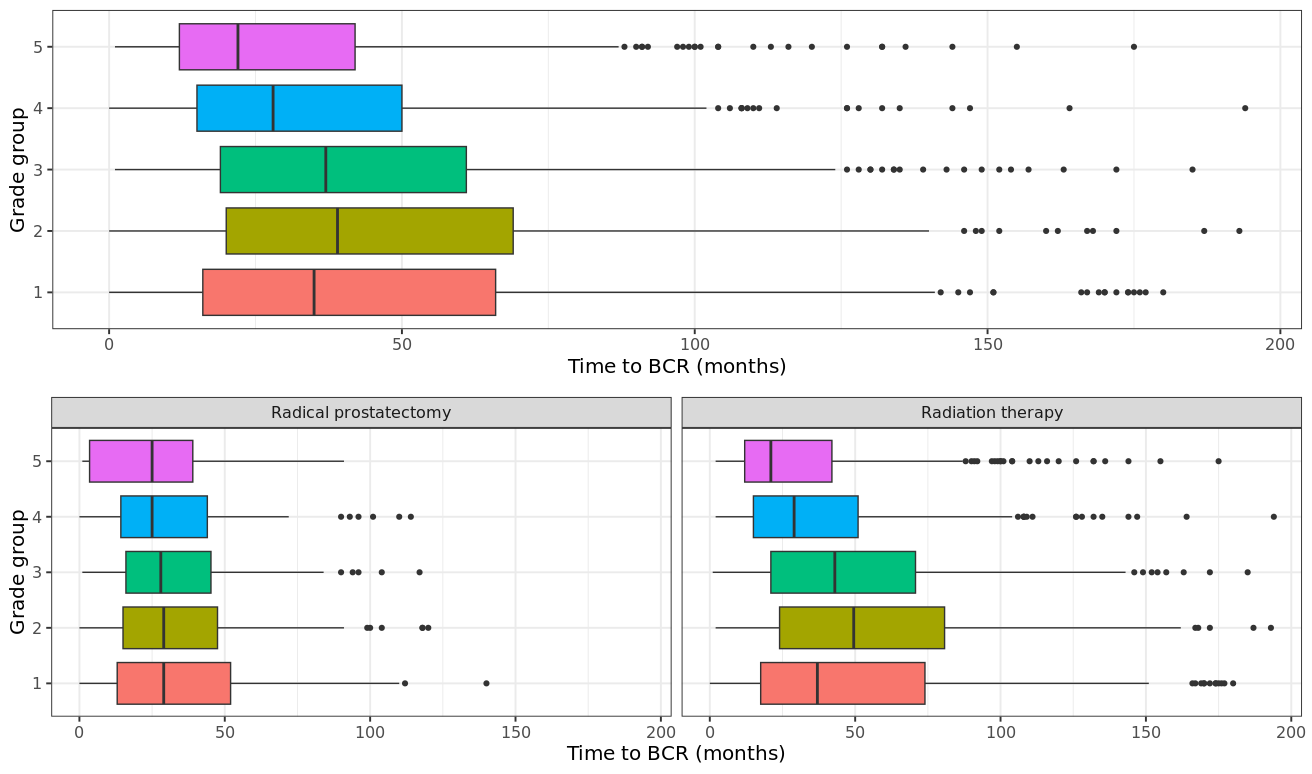}
    \caption{BCR detected data: The time to relapse by Gleason grade group.}
    \label{fig:bcr_ggs}
\end{figure}

Our DBCR algorithm successfully identified 2851 patients (Figure \ref{fig:bcr_dist}-\ref{fig:bcr_ggs}) who developed a biochemical recurrence after a curative treatment (27\% of the treated patients). Among them, 70\% (n=2007) were detected using the PRP and PRT methods in accordance with the EAU guidelines \cite{van_den_broeck_biochemical_2020}, while 30\% (n=844) were identified using our new algorithms (CRP, CRT), which might have otherwise gone unnoticed.

\subsection{Applications}
\begin{figure}
    \centering
    \includegraphics[width=1\linewidth]{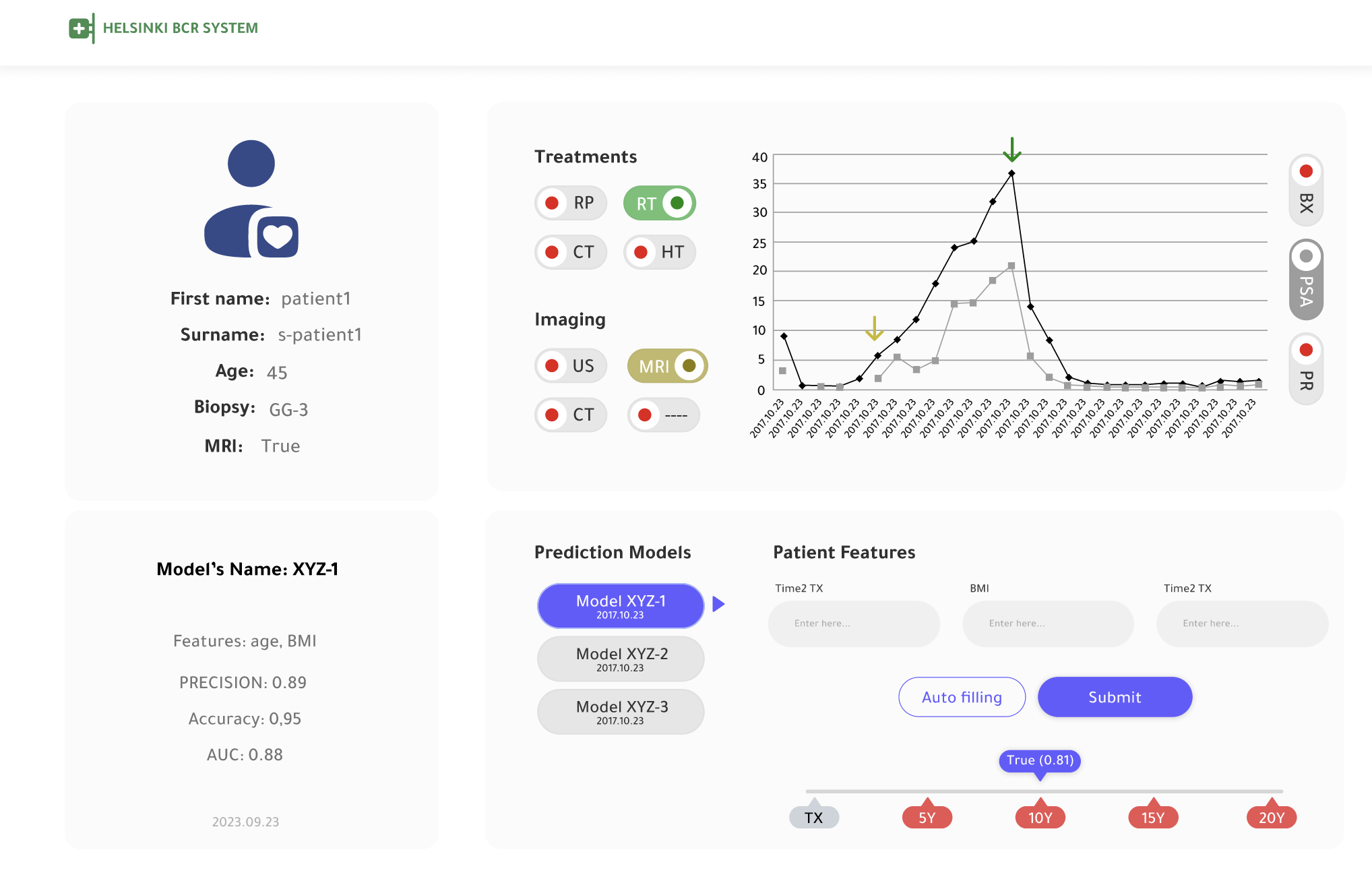}
    \caption{Helsinki BCR system enabling access to patients' trajectory and the BCR prediction models.}
    \label{fig:bcr_system}
\end{figure}

After cleaning the data and improving its quality, we have successfully carried out multiple applications. The primary application involved developing a data visualisation tool, enabling clinicians and researchers to visualise the trajectory of PCa patients, including their PSA values, treatments, pathological results, medical prescriptions, and others (Figure \ref{fig:bcr_system}).

Additionally, we have investigated the potential grade inflation in PCa Gleason grade groups during the MRI era; the research focused on patients in Gleason grade groups 1 and 2. The hypothesis of this work was that some patients in grade group 1 in the pre-MRI era are nowadays, in the MRI era, classified and treated as grade group 2 patients. With enough evidence, the work proved the hypothesis which will open serious discussions to reassess current risk stratification tools and clinical decision-making. Updating guidelines on cancer grading and treatments is crucial to be aligned with the precision of modern MRI technology.

Furthermore, we are utilising the curated EHR data to train machine learning models to predict biochemical recurrence within the following 3-to-10 years from initial curative treatment. Knowing that prostate cancer is a slow-developing cancer, BCR is one of the most important and accurate surrogates to prostate cancer mortality. Therefore, predicting BCR would have a significant impact on treatment decisions and treatment planning. Our (preliminary) trained models achieved good performance (Accuracy=0.93, AUC=0.93, Precision=0.88) on an internal validation. The models are trained on n=5262 patients who have had PCa curative treatment.

\FloatBarrier

\section{\uppercase{Discussion}}
Ensuring high data quality is essential when building effective AI models and conducting significant statistical analyses. This importance is particularly heightened in clinical research and applications where decisions may directly impact patients' lives. Electronic Health Records (EHR), like HUS Datalake, play a critical role in this process, making it imperative to develop robust exploration methods to harness the available data. 

In our work, we explored, curated, and augmented bio-medical data from within Finnish healthcare records, with a specific focus on prostate cancer patients. By establishing a new mining framework and developing novel analysis algorithms, we successfully consolidated our data, enabling us to conduct meaningful and impactful medical research. One of our approaches was to use the time series data on patients' PSA levels, a subset of medical data which is typically well collected and curated within EHR, in order to infer the existence, and the type, of EHR missing curative treatment events. To our knowledge, this is the first time PSA time series data were used in this way, although, in \cite{joa15}, the authors employed a similar approach in order to generate a completeness score for the overall data quality of the cohort. Based on this approach, we were able to consolidate our EHR by adding approx. $2.8$k new curative treatment events, representing a $27\%$ increase from the EHR-available treatment events. 

Another important outcome of our mining framework was documenting the status and timing of our PCa patients' BCR. Differently than in previous EHR mining frameworks for PCa medical data, see e.g. \cite{park21,par21}, we define BCR-status based on both PSA-level measurements (after primary curative treatment, i.e., radiation therapy --RT-- or radical prostatectomy --RP--) as well as based on secondary curative and adjuvant therapies, i.e., PCa related hormonal- and/or chemotherapy. This approach takes into consideration the clinical reality that sometimes, curating doctors decide on secondary therapies before the PSA level crosses the threshold established by current EAU guidelines as the BCR level. Using this approach, we accurately captured an additional 844 BCR events (representing a $42\%$ increase from PSA-only detected BCR events), which otherwise would either not have been found at all or would have been given a significant later time-stamp.  

One important observation from our EHR data curation and analysis work is that there exists a large amount of redundancy in these data sources. This is particularly observable within the free text input written by doctors during their medical checkups and/or lab, pathological, or imaging reports. On the other hand, due to a multitude of factors, including human error, focusing on only one particular type of data source at a time, such as lab results, pathological reports, or even surgery records, one encounters a significant amount of missing data entries. Thus it becomes both feasible and greatly beneficial to use the data redundancy feature of EHR in order to "recuperate" these missing data entries. This is why, a "data investigation" approach, such as the one described in this manuscript, is more relevant than classical "data imputation" methods. Indeed, these latter approaches provide only average-like behaviours and also are completely inefficient in detecting missing events, such as a radiation treatment event altogether missing from within the EHR.

Strongly connected to the above reasoning, one could not overlook the potential impact the use of Large Language Models (LLM) could have in detecting and augmenting the existing EHR data \cite{llm_2023}. Such models could be employed to extract (from the free text provided by doctors) relevant information such as missing events, e.g. treatments performed in different clinics, cities, or even countries, or information that is usually not structurally recorded within EHR, e.g., family history, use of alcohol and tobacco products, general health status of the patient, etc. During the current EHR data analysis no LLM was employed; however, the approach is currently actively analysed for future usage within our models.

\section{\uppercase{Conclusion}}
This work demonstrates the challenges of mining Finnish electronic health records for prostate cancer (PCa) research, as well as the opportunities it offers in gaining valuable insights. Our methodology, when applied to the HUS datalake, enabled the detection of missing treatments and biochemical recurrences (BCR), which led to a range of clinically relevant findings, including patients' timeline histories, the Gleason grade group inflation finding, and the BCR classification models. The results of our framework highlight the potential of EHR data mining to advance PCa research and guide personalised patient care.

\section*{\uppercase{Acknowledgements}}

This work was supported by grants from the Cancer Society Finland, the Academy of Finland, Jane and Aatos Erkko Foundation, and State funding for university-level health research. It is a joint effort of doctoral students, senior researchers, and clinicians at the University of Helsinki and the University Hospital of Helsinki.

\bibliographystyle{apalike}
{\small
\bibliography{main}}

\section*{\uppercase{Appendix}}

\begin{algorithm}[h]
 \caption{DTX - Missing treatments detection}
 \label{algo_dtx}
 \begin{algorithmic}[1]
 \renewcommand{\algorithmicrequire}{\textbf{Input:}}
 \renewcommand{\algorithmicensure}{\textbf{Output:}}
 \renewcommand{\algorithmiccomment}[1]{\hfill$\triangleright$\textit{#1}}
 
 \REQUIRE $PATIENTS\_LIST$
 \ENSURE  $L$
 
  \STATE $L \leftarrow [\;]$
  \FORALL{$p_i$ \textbf{in} $PATIENTS\_LIST$}
   \STATE $PSA_i \leftarrow getPsa(p_i)$
   \STATE $Tx_i \leftarrow getTreatments(p_i)$
   \STATE $(d_{max}, d_{min}, PSA_{min}) \leftarrow SIGDROP(PSA_i)$
   
   \IF{$TxExisits(d_{max}, d_{min}, PSA_{min}, Tx_i) = False$}
        \IF{$PSA_{min} < 0.1$}
            \STATE $tx\_type \leftarrow $ 'RP'
        \ELSE
            \STATE $tx\_type \leftarrow $ 'RT'
        \ENDIF
        \STATE $L \leftarrow L + (p_i, tx\_type, drop\_date)$
   \ENDIF
  \ENDFOR
 \RETURN $L$
 \end{algorithmic}
\end{algorithm}

\begin{algorithm}[!ht]
 \caption{PRP - PSA Relapse after RP}
 \label{algo_prp}
 \begin{algorithmic}[1]
 \renewcommand{\algorithmicrequire}{\textbf{Input:}}
 \renewcommand{\algorithmicensure}{\textbf{Output:}}
 \renewcommand{\algorithmiccomment}[1]{\hfill$\triangleright$\textit{#1}}
 
 \REQUIRE $p_i$
 \ENSURE  $d_{m2}$
   \STATE $PSA \leftarrow getPsaAfterRp(p_i)$
   \FOR{$psa_j$ \textbf{in} $PSA$}
       \IF{$usp(psa_j) = TRUE$}
            \STATE $th \leftarrow 0.2$ 
        \ELSE
            \STATE $th \leftarrow 0.4$ 
       \ENDIF
        \IF{$psa_j > th$}
            \STATE $d_{m2} \leftarrow getDate(psa_j)$
            \RETURN $d_{m2}$
        \ENDIF
  \ENDFOR
 \RETURN NULL
 \end{algorithmic}
\end{algorithm}

\begin{algorithm}[!ht]
 \caption{PRT - PSA Relapse after RT}
 \label{algo_prt}
 \begin{algorithmic}[1]
 \renewcommand{\algorithmicrequire}{\textbf{Input:}}
 \renewcommand{\algorithmicensure}{\textbf{Output:}}
 \renewcommand{\algorithmiccomment}[1]{\hfill$\triangleright$\textit{#1}}
 
 \REQUIRE $p_i$
 \ENSURE  $d_{m2}$
 
   \STATE $PSA \leftarrow getPsaAfterRt(p_i)$
   \STATE $nadir \leftarrow getMax(PSA)$
   \FOR{$psa_j$ \textbf{in} $PSA$}
        \IF{$nadir > psa_j$}
            \STATE $nadir \leftarrow psa_j$
        \ENDIF
        \STATE $inc \leftarrow psa_j - nadir$
        \IF{$inc > 2$}
            \STATE $d_{m2} \leftarrow getDate(psa_j)$
            \RETURN $d_{m2}$
        \ENDIF
  \ENDFOR

 \RETURN $NULL$
 \end{algorithmic}
\end{algorithm}

\begin{algorithm}[ht]
 \caption{DBCR}
 \label{algo_dbcr_main}
 \begin{algorithmic}[1]
 \renewcommand{\algorithmicrequire}{\textbf{Input:}}
 \renewcommand{\algorithmicensure}{\textbf{Output:}}
 \renewcommand{\algorithmiccomment}[1]{\hfill$\triangleright$\textit{#1}}
 
 \REQUIRE $TREATED\_PATIENTS$
 \ENSURE  $L_{bcr}$
 
  \STATE $L_{bcr} \leftarrow [\;]$
  \FORALL{$p_i$ \textbf{in} $TREATED\_PATIENTS$}

   \STATE $d_{1} \leftarrow PRP(p_i)$
   \STATE $d_{2} \leftarrow CRP(p_i)$
   \STATE $d_{3} \leftarrow PRT(p_i)$
   \STATE $d_{4} \leftarrow CRT(p_i)$

   \IF{$allAreNULL(d_{1},d_{2},d_{3},d_{4} ) = FALSE$}       
        \STATE $bcr\_date \leftarrow getMin(d_{1},d_{2},d_{3},d_{4})$
        \STATE $new\_bcr \leftarrow (p_i, bcr\_date)$
        \STATE $L_{bcr} \leftarrow L_{bcr} + new\_bcr$
   \ENDIF
   
  \ENDFOR
 \RETURN $L_{bcr}$
 \end{algorithmic}
\end{algorithm}

\end{document}